\documentclass[aps,prl,floatfix,twocolumn,superscriptaddress]{revtex4-2}
\usepackage[normalem]{ulem}
\usepackage{bm}
\usepackage{color}
\usepackage{physics}

\usepackage{graphicx,
            caption,
            subcaption,
            amsmath,
            braket,
            units,
            ragged2e,
            xcolor,
            xspace,
            nicefrac,
            lipsum,
            nameref,
            textcomp,
            urwchancal,
            upgreek,
            isotope,
            hyperref,
            cleveref,
            siunitx,
            placeins, 
            scrextend, 
            url,
            float}

\DeclareCaptionLabelSeparator{dot}{. }
\makeatletter
\def\justified{
	\let\\\@normalcr
	\@rightskip\z@skip \rightskip\@rightskip
	\leftskip\z@skip
	\parindent 0em\relax
	\setlength{\parfillskip}{0pt plus 1fil}}
\DeclareCaptionJustification{justified}{\justified}
\hypersetup{colorlinks=true, linkcolor=blue,citecolor=blue,urlcolor=blue}
\newcommand{\myref}[2][]{Fig.~\hyperref[#2]{\ref*{#2}#1}}
\newcommand{\Myref}[2][]{Figure~\hyperref[#2]{\ref*{#2}#1}}
\newcommand{\Mytabref}[2][]{Table~\hyperref[#2]{\ref*{#2}#1}}

\newcommand{\edd}{\epsilon_\text{dd}}

\setcitestyle{super}

\begin{document}

\title{Observation of vortices and vortex stripes in a dipolar Bose-Einstein condensate}

 \author{Lauritz Klaus}
 \thanks{These two authors contributed equally to this work.}
 \author{Thomas Bland}
 \thanks{These two authors contributed equally to this work.}
  \affiliation{
     Institut f\"{u}r Quantenoptik und Quanteninformation, \"Osterreichische Akademie der Wissenschaften, Innsbruck, Austria}
 \affiliation{
     Institut f\"{u}r Experimentalphysik, Universit\"{a}t Innsbruck, Austria }
 
 \author{Elena Poli}
  \affiliation{
     Institut f\"{u}r Experimentalphysik, Universit\"{a}t Innsbruck, Austria }
     
 \author{Claudia Politi}
 \affiliation{
     Institut f\"{u}r Quantenoptik und Quanteninformation, \"Osterreichische Akademie der Wissenschaften, Innsbruck, Austria}
 \affiliation{
     Institut f\"{u}r Experimentalphysik, Universit\"{a}t Innsbruck, Austria }

 \author{Giacomo Lamporesi}
  \affiliation{INO-CNR BEC Center and Dipartimento di Fisica, Universit\`a di Trento, 38123, Trento, Italy}
 
 \author{Eva Casotti}
  \affiliation{
     Institut f\"{u}r Quantenoptik und Quanteninformation, \"Osterreichische Akademie der Wissenschaften, Innsbruck, Austria}
 \affiliation{
     Institut f\"{u}r Experimentalphysik, Universit\"{a}t Innsbruck, Austria }

 \author{Russell N. Bisset}
   \affiliation{
     Institut f\"{u}r Experimentalphysik, Universit\"{a}t Innsbruck, Austria }
     
 \author{Manfred J. Mark}
 \author{Francesca Ferlaino}
  \thanks{Correspondence should be addressed to \mbox{\url{Francesca.Ferlaino@uibk.ac.at}}}
  \affiliation{
     Institut f\"{u}r Quantenoptik und Quanteninformation, \"Osterreichische Akademie der Wissenschaften, Innsbruck, Austria}
 \affiliation{
     Institut f\"{u}r Experimentalphysik, Universit\"{a}t Innsbruck, Austria }

\begin{abstract}
Quantized vortices are the prototypical feature of superfluidity. Pervasive in all natural systems, vortices are yet to be observed in dipolar quantum gases.
Here, we exploit the anisotropic nature of the dipole-dipole interaction of a dysprosium Bose-Einstein condensate to induce angular symmetry breaking in an otherwise cylindrically symmetric pancake-shaped trap. Tilting the magnetic field towards the radial plane deforms the cloud into an ellipsoid through magnetostriction, which is then set into rotation.
At stirring frequencies approaching the radial trap frequency, we observe the generation of dynamically unstable surface excitations, which cause angular momentum to be pumped into the system through vortices.
Under continuous rotation, the vortices arrange into a stripe configuration along the field--in close corroboration with simulations--realizing a long sought-after prediction for dipolar vortices.
\end{abstract}

\maketitle

Since the first experiments on gaseous Bose-Einstein condensates (BECs), the observation of quantized vortices has been considered the most fundamental and defining signature of the superfluid nature of such systems. Their very existence sets a unifying concept encompassing a variety of quantum fluids from liquid helium \cite{Donnelly1991qvi} to the core of neutron stars \cite{Pines1985sin}, and from superconductors \cite{Abrikosov2004nlt} to quantum fluids of light \cite{Lagoudakis2008qvi}. Their classical counterparts have as well fascinated scientists from different epochs and fields as vortices are found in many scales of physical systems, from tornadoes in the atmosphere to ferrohydrodynamics.

In the quantum realm, a quantized vortex may emerge as a unique response of a superfluid to rotation. It can be understood as a type of topologically protected singularity with a $2\pi$ phase winding that preserves the single-valuedness of the superfluid wavefunction and the irrotational nature of its velocity field. In contact-interacting BECs, vortical singularities have been observed experimentally in the form of single vortices \cite{Matthews1999via,Madison2000vfi}, vortex-antivortex pairs \cite{neely2010observation}, solitonic vortices \cite{Ku2014moa,Donadello2014oos}, vortex rings \cite{Anderson2001wds} and vortex lattices \cite{Madison2000vfi,Abo-Shaeer2001oov, Coddington2003oot} using a number of different techniques.
Moreover, vortices play a fundamental role in the description of the Berezinskii-Kosterlitz-Thouless transition in two-dimensional systems \cite{Hadzibabic2006bkt}, as well as in the evolution of quantum turbulence \cite{Neely2013cot,Navon2016eoa, Tsatsos2016qti}, and have been observed in interacting Fermi gases along the BEC-to-BCS crossover \cite{Zwierlein2005vas,Ku2014moa}. 

Recently, a new class of ultracold quantum gases are being created in various laboratories around the world, using strongly magnetic lanthanide atoms \cite{Lu2011sdb, Aikawa2012bec}. 
Such a system, providing a quantum analogue of classical ferrofluids, enables access to the physics of dipolar BECs, in which atoms feature a strong long-range anisotropic dipole-dipole interaction (DDI) \cite{Norcia2021dia,Chomaz2022dpa} on top of the traditional contact-type isotropic one.  
This intriguing platform provided the key to observe, e.g., extended Bose-Hubbard dynamics \cite{Baier2016ebh}, roton excitations \cite{Landau1941hto,Chomaz2018oor, Schmidt2021rei}, the quantum version of the Rosensweig instability \cite{Kadau2016otr}, supersolid states of matter \cite{Tanzi2019ooa,Boettcher2019tsp,Chomaz2019lla, Norcia2021tds}, and is foreseen to host novel phenomena for quantum simulation and metrology \cite{Norcia2021dia,Chomaz2022dpa}.

The dipolar interaction is predicted to also intimately change the properties of vortices in quantum gases \cite{Martin2017vav}. For instance, theoretical works predict single vortices to exhibit an elliptic-shaped core for a quasi-two-dimensional setting with in-plane dipole orientation \cite{yi2006vsi,Ticknor2011asi,mulkerin2013anisotropic,mulkerin2014vortices}, or the presence of density oscillations around the vortex core induced by the roton minimum in the dispersion relation \cite{yi2006vsi,Ticknor2011asi,mulkerin2013anisotropic,mulkerin2014vortices,JonaLasinio2013rci}. For vortex pairs, the anisotropic DDI is expected to alter the lifetime and dynamics \cite{mulkerin2014vortices,gautam2014dynamics}, and can even suppress vortex-antivortex annihilation \cite{mulkerin2014vortices}. These interaction properties are predicted to give rise to a vortex lattice structure that can follow a triangular pattern \cite{yi2006vsi,JonaLasinio2013rci}, as is typical for non-dipolar BECs \cite{Abo-Shaeer2001oov}, or a square lattice for attractive or zero contact interactions \cite{Cooper2005vli,Kumar2016tdv} when the DDI is isotropic (dipoles aligned with the rotation axis). A very striking consequence of the dipoles tilted into the plane is the formation of vortex stripes \cite{yi2006vsi,cai2018vpa,Prasad2019vlf}.
Moreover, vortices could provide an unambiguous smoking gun of superfluidity in
supersolid states \cite{Roccuzzo2020ras,Gallemi2020qvi, Ancilotto2021vpi, Norcia2021cao, Roccuzzo2022moi}.
However, despite these intriguing predictions, vortices in dipolar quantum gases have yet to be observed.

This paper presents the experimental realization of quantized vortices in a dipolar BEC of highly magnetic dysprosium atoms. Following a method proposed in Ref.\,\cite{Prasad2019vlf}, we show that the many-body phenomenon of magnetostriction \cite{Stuhler2005ood}, genuinely arising from the anisotropic DDI among atoms, provides a natural route to rotate the systems and nucleate vortices in a dipolar BEC. We carry out studies on the dynamics of the vortex formation, which agree very well with our theoretical predictions. Finally, we observe one of the earliest predictions for dipolar vortices: the formation of vortex stripes in the system.

In non-dipolar gases, quantized vortices have been produced using several conceptually different techniques. For instance by rotating non-symmetric optical \cite{Madison2000vfi,Abo-Shaeer2001oov} or magnetic \cite{Haljan2001dbe} potentials, by rapidly shaking the gas \cite{Navon2016eoa}, by traversing it with obstacles with large enough velocity \cite{neely2010observation,Kwon2021sea}, by rapidly cooling the gas across the BEC phase transition \cite{Corman2014qis, Liu2018dea}, or directly imprinting the vortex phase pattern \cite{delPace2022ipc}.
Dipolar quantum gases, while able to form vortices with these same standard procedures \cite{Martin2017vav}, also offer unique opportunities that have no counterpart in contact-interacting gases.
Crucially, the DDI gives rise to the phenomenon of magnetostriction in position space \cite{Stuhler2005ood}. When dipolar BECs are polarized by an external magnetic field--defining the dipole orientation--the DDI causes an elongation of the cloud along the polarization direction. This is a direct consequence of the system tendency to favor head-to-tail dipole configurations, which effectively reduces the interaction energy \cite{Chomaz2022dpa}. 

\begin{figure}[t!]
\includegraphics[width=0.85\columnwidth]{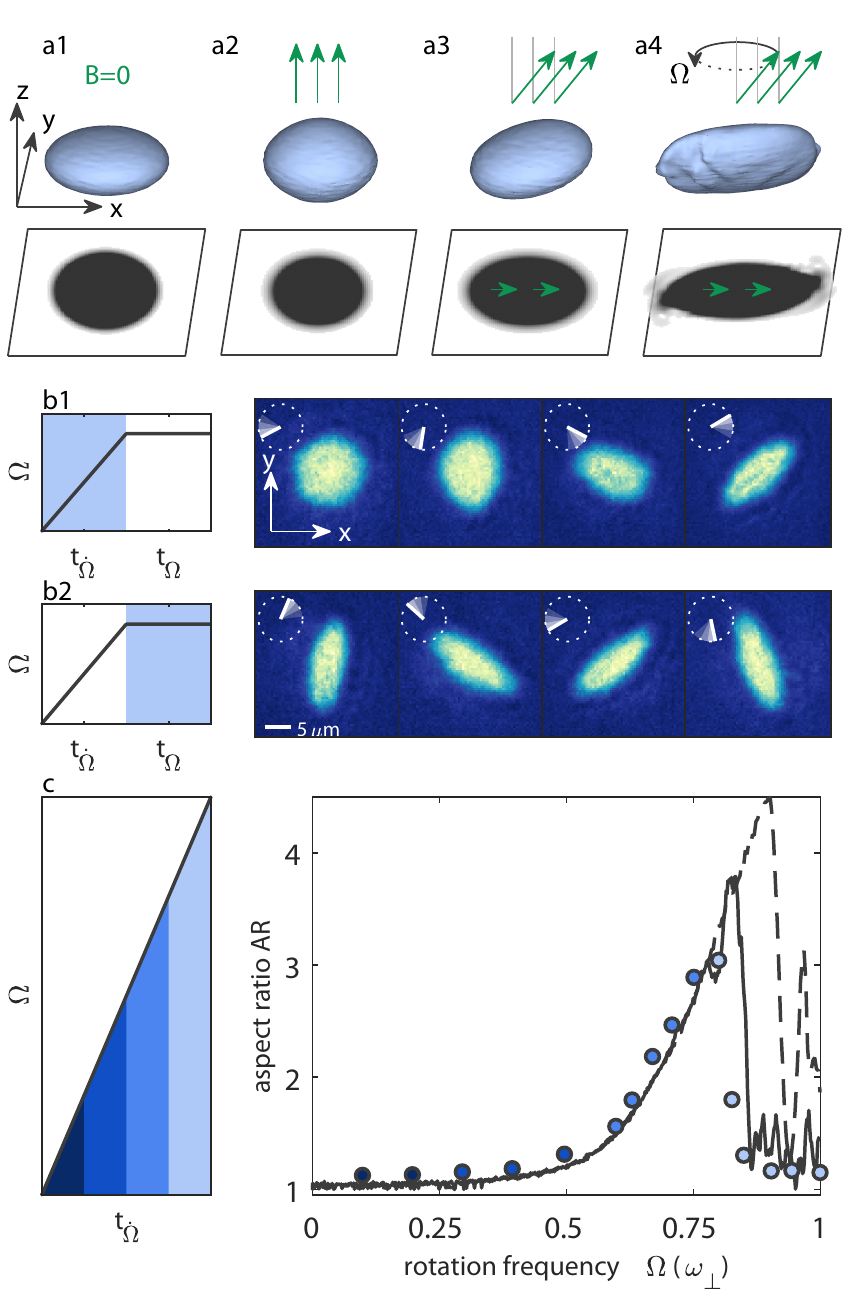}
\caption{\textbf{Magnetostirring of a Dy dipolar BEC and evolution of the cloud aspect ratio.} \textbf{a}, Illustration of a non-dipolar (\textbf{a1}) and dipolar BEC with ${\rm B}\neq 0$ (\textbf{a2-a4}) in a cylindrically symmetric, oblate trap. 
The magnetic-field angle with respect to the $z$ axis varies from $\theta=0$° (\textbf{a2}) to $\theta = 35^\circ$ (\textbf{a3}) and  rotating at $\theta = 35^\circ$ around $z$ (\textbf{a4}). \textbf{b}, Representative axial absorption images showing the dipolar BEC  while spinning up the magnetic field for $t_{\dot{\Omega}} = [140, 430, 627, 692]\,\si{ms}$ (\textbf{b1})  and  subsequent constant rotation at $\Omega =2\pi \times 36\,\si{Hz}$ for $t_\Omega = [0, 6, 11, 17]\,\si{ms}$ (\textbf{b2}). 
\textbf{c}, Cloud aspect ratio AR for different final rotation frequencies. $\Omega$ is linearly increased to its final value at a speed of $\dot{\Omega} = 2\pi \times 50\, \si{Hz}/\si{s}$. To mitigate influences of trap anisotropies on the AR, a full period at the final rotation frequency is probed. The error bars are smaller than the symbol size. The solid (dashed) black line shows the corresponding eGPE simulations with a $2\,\si{s}$ ($1\,\si{s}$) ramp and $a_s = 110\, a_0$, $(\omega_\perp, \omega_z) = 2\pi \times [50,130]\,\si{Hz}$, and $N=15000$. } 
\label{fig:Fig1}
\end{figure}

Such a magnetostrictive effect provides a simple method to induce an elliptic effective potential and drive rotation with a single control parameter. This modification of the effective potential is shown in Fig.\,\ref{fig:Fig1}a for a BEC in an oblate trap with cylindrical symmetry about the $z$-axis.
While a non-dipolar BEC takes the same shape as the confining trap (a1), 
introducing dipolar interactions with polarization axis along $z$ stretches the cloud along this axis, yet maintains cylindrical symmetry (a2). Tilting the ${\va*{\rm B}}$-field leads to a breaking of the cylindrical symmetry, resulting in an ellipsoidal deformation of the cloud shape, as seen from the density projection onto the $x$-$y$ plane  (a3).
Finally, under continuous rotation of the magnetic-field, which we coin ``magnetostirring", the condensate is predicted to rotate (a4).  
This unique approach to stir a dipolar condensate can eventually lead to the nucleation of vortices \cite{Prasad2019vlf,Prasad2021aar}, realizing genuinely interaction-driven vorticity through many-body phenomena. 

\begin{figure*}
\includegraphics[width=2\columnwidth]{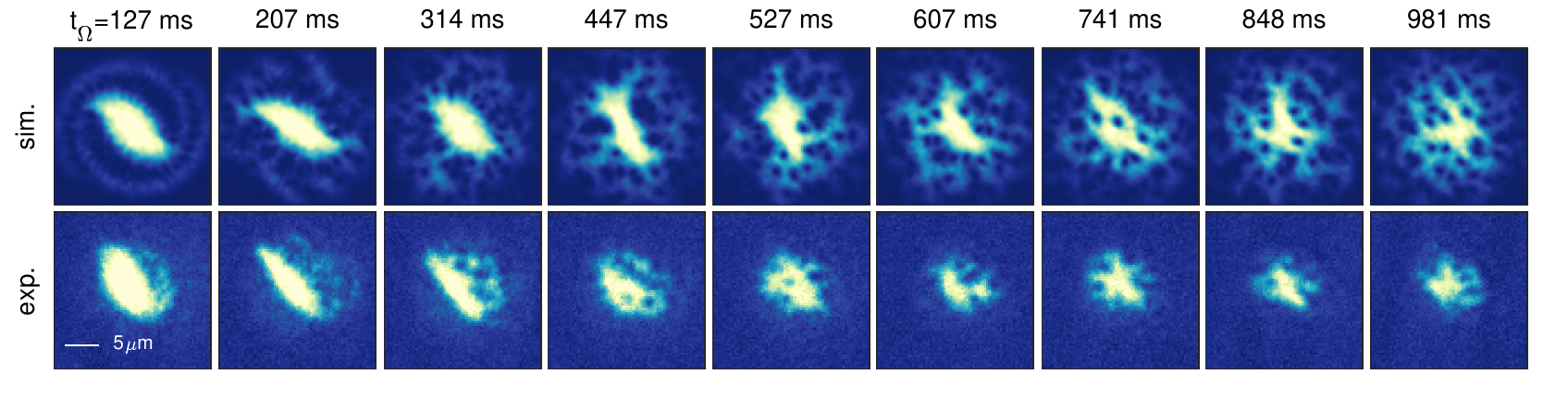}
\caption{\textbf{Observation of vortices in a dipolar BEC.} Each column shows the simulated (upper) and experimental (lower) images for various rotation times $t_{\Omega}$. For the experiment, the atoms are imaged along the $z$ direction.  In each experimental run, we rotate the magnetic field counter-clockwise at $\Omega = 0.74\,\omega_\perp$ for different rotation times $t_{\Omega}$.
The magnetic field value is kept to $|{\va*{\rm B}}|=5.333(5)\,\si{G}$.
The initial condensed atom number is $N=15000$. The decreasing size of the cloud suggests a decrease in atom number. However, for states with vortices or spiral shapes, appearing at large $t_\Omega$, our bimodal fit to extract the atom number breaks down. 
For the corresponding simulations, the parameters are  $a_s=112\,\si{a_0}$, trap frequencies $(\omega_\perp, \omega_z)=2\pi\times[50, 150]\,\si{Hz}$, $N=8000$, and  $\Omega=0.75\,\omega_\perp$.}
\label{fig:Fig2}
\end{figure*}

We explore this protocol using a dipolar BEC of $^{162}$Dy atoms. We create the BEC similarly to our previous work \cite{Norcia2021cao} with the distinction that here the magnetic field unit vector, $\vu*{\rm B}$, is kept tilted at an angle of $\theta = 35$° with respect to the $z$-axis both during evaporative cooling and magnetostirring (see Fig.\,\ref{fig:Fig1}a3 and Methods).
After preparation, the sample contains about $2 \times 10^4$ condensed atoms confined within a cylindrically symmetric optical dipole trap (ODT) with typical radial and axial trap frequencies $(\omega_\perp, \omega_z) = 2\pi\times[50.8(2), 140(1)]\,\si{Hz}$. 
Here, prior to stirring, the magnetostriction is expected from simulations to increase the cloud aspect ratio in the horizontal plane from 1 up to 1.03, whereas the trap anisotropy is negligible.
We use a vertical ($z$) absorption imaging to probe the radial ($x$,$y$) atomic distribution after a short time-of-flight (TOF) expansion of $3\,\si{ms}$. The atom number is instead measured using horizontal absorption imaging with a TOF of $26\,\si{ms}$.

Similarly to a rotation of a bucket containing superfluid helium or of a smoothly deformed ODT for non-dipolar BECs, magnetostirring is predicted to transfer angular momentum into a dipolar BEC \cite{Prasad2019vlf,Prasad2021aar}. In response to such an imposed rotation, the shape of an irrotational cloud is expected to elongate with an amplitude that increases with the rotation frequency $\Omega$.  This phenomenon is clearly visible in our experiments, as shown in Fig.\,\ref{fig:Fig1}b. 
Here, we first revolve the tilted $\vu*{\rm B}$ around the $z$-axis with a linearly increasing rotation frequency ($\dot{\Omega} = 2\pi \times 50\,\si{ Hz/s}$) and observe that the dipolar BEC starts to rotate at the same angular speed as the field and deforms with increasing elongation (b1). We then stop the adiabatic ramp at a given value of $\Omega$ and probe the system under continuous rotation. We now find that the cloud continues rotating in the radial plane with an almost constant shape (b2). Note that $|{\va*{\rm B}}|$  is held constant at $5.333(5)\,\si{G}$, where we estimate a contact scattering length of about $a_s=111\,a_0$ (Methods).

We further explore the response of our dipolar BEC to magnetostirring by repeating the measurements in Fig.\,\ref{fig:Fig1}b1, but stopping the ramp at different final values of $\Omega$. The maximum value used for $\Omega$ approaches $\omega_\perp$, corresponding to a ramp duration of $1\,\si{s}$.
We quantify the cloud elongation in terms of the aspect ratio AR$=\sigma_\text{max}/\sigma_\text{min}$, where the cloud widths $\sigma_\text{max}$ and  $\sigma_\text{min}$ are extracted by fitting a rotated two-dimensional Gaussian function to the density profiles. Figure \ref{fig:Fig1}c summarizes our results. We observe that initially the AR slightly deviates from one due to magnetostriction. It then slowly grows with increasing $\Omega$, until a rapid increase at around 0.6$\,\omega_\perp$ occurs, as this allows the angular momentum to increase, which decreases the energy in the rotating frame\cite{Recati2001oov}.
Suddenly, at $\Omega_c \approx 0.74 \,\omega_\perp$, the AR abruptly collapses back to $\mathrm{AR} \approx 1$, showing how the superfluid irrotational nature competes with the imposed rotation.

To substantiate our observation, we perform numerical simulations of the zero-temperature extended Gross-Pitaevskii equation (eGPE) \cite{Waechtler2016qfi, FerrierBarbut2016ooq, Chomaz2016qfd, Bisset2016gsp} (see Methods). Quantum and thermal fluctuations are added to the initial states, which are important to seed the dynamic instabilities once they emerge at large enough $\Omega$; see later discussion.
The lines in Fig.\,\ref{fig:Fig1}c show our results. The dashed line is obtained through the same procedure as the experiment, whereas for the solid line we halve the ramp rate, spending more time at each frequency. Both ramp procedures show quantitatively the same behaviour up to $\Omega = 0.8\,\omega_\perp$, and are in excellent agreement with the experimental results. The stability of the $1\si{s}$ ramp exceeds the experimentally observed critical frequency. We partly attribute this discrepancy to asymmetries of the rotation in the experiment that are not present in the simulations, which may lead to an effective speedup of the dynamical instabilities. However, in all cases, the AR rapidly decreases to about one.

The growing AR and subsequent collapse to one is a signature of the dynamical instability of surface modes, known for being an important mechanism for seeding vortices and allowing them to penetrate into the high-density regions of rotated BECs \cite{Recati2001oov, Sinha2001dio, Madison2001sso}, as also predicted for our dipolar system \cite{Prasad2019vlf}.
To search for quantum vortices in our system, we perform a new investigation where we directly set $\Omega$ close to $\Omega_c$. We then hold the magnetic-field rotation fixed at this constant frequency for a time $t_\Omega$. As shown in Fig.\,\ref{fig:Fig2} (lower row), the cloud rapidly elongates, and the density starts to exhibit a spiral pattern, emanating from the tips of the ellipsoid. As early as $t_\Omega=314\,\si{ms}$ clear holes are observed in the density profile, forming in the density halo around the center, the first clear indication of vortices in a dipolar gas. These vortices, initially nucleated at the edge of the sample, persist as we continue to stir, and eventually migrate towards the central (high-density) region. Owing to their topological protection, vortices are still visible in the experiment after one second of magnetostirring, although our atom number decreases throughout this procedure.
Our observations bear a remarkable resemblance to the simulations; Fig.\,\ref{fig:Fig2} (upper row) showing the in situ column densities. Taking a fixed atom number of $N=8000$, but otherwise repeating the experimental sequence, we observe many similar features. First, the spiral density pattern appears before the instability, forming two arms that are filled with vortices close to the central density. Next, turbulent dynamics ensue as the density surface goes unstable and vortices emerge in the central high density region. This turbulence, however, inhibits the creation of a vortex lattice on these timescales.

\begin{figure}
\includegraphics[width=3.38 in, ]{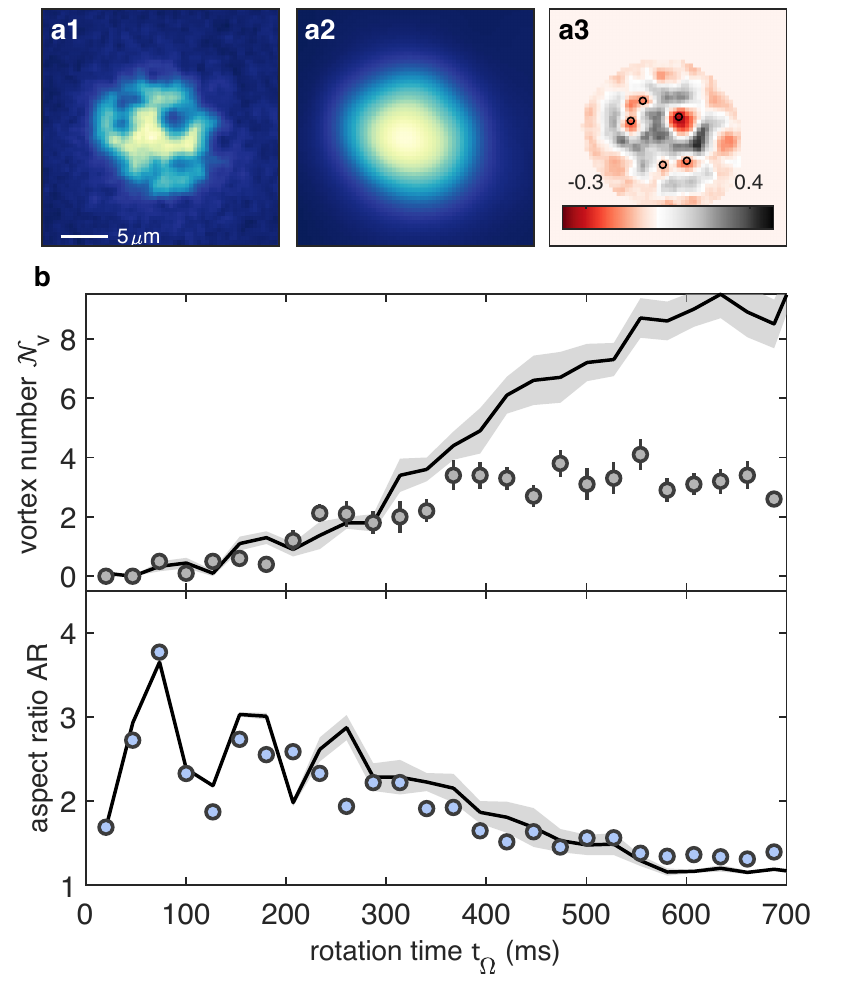}
\caption{\textbf{Time evolution of the average vortex number, $\mathcal{N}_v$, and cloud aspect ratio AR.} \textbf{a1}, Sample image after rotating for $t_\Omega = 474\,\si{ms}$. \textbf{a2}, Blurred reference image ($\sigma=2.1\,\mu$m). \textbf{a3}, Residuals with markers (black circles) indicating the identified vortices. \textbf{b}, The detected vortex number $\mathcal{N}_v$ (top panel) and the AR of the cloud (bottom panel) after the rotation time $t_\Omega$. Data points and error bars show the mean and standard error from about 10 experimental runs.  Solid lines indicate the averaged results from 10 corresponding simulations with different initial noise for parameters $a_s = 110\, a_0$, $(\omega_{\perp}, \omega_z) = 2\pi \times [50,130]$\,Hz, $N=10000$ and $\Omega = 0.75\,\omega_{\perp}$, the shaded area gives its standard error.}
\label{fig:Fig3}
\end{figure}

The observed evolution of the system under constant rotation shows some concurrence between the appearance of vortices in the absorption images and the formation of a round density pattern in the radial plane with $\mathrm{AR}\approx 1$ (Fig.\,\ref{fig:Fig2}). To study this dynamical evolution in more detail, we adopt an analysis protocol for both the experiment and theory that allows us to quantitatively track the evolution of the average number of vortices, $\mathcal{N}_v$ (Methods). The result is shown in Fig.\,\ref{fig:Fig3}a. 
In brief, for each single image (a1) we create a blurred reference image by applying a 2D Gaussian filter \cite{Abo-Shaeer2002fad,Kwon2014ros}. 
We then calculate the difference between each single image (a1) and the corresponding reference (a2) to obtain the residual image (a3), from which we count $\mathcal{N}_v$ by finding local minima below a certain threshold. 
 
For the experimental density profiles, which are affected by both the limited resolution of the imaging system and the weak contrast in the low-density zones (halo) where the vortices initially nest, we expect $\mathcal{N}_v$ to be underestimated relative to the number expected by theory. 
However, in order to carry out a quantitative comparison with the simulations, we apply a blurring filter and add noise to the latter that mimics the actual resolution in the experiment (Methods). 

Figure \ref{fig:Fig3}b shows both the evolution of $\mathcal{N}_v$ and cloud AR as a function of rotation time, $t_\Omega$. Solid lines are the results from the eGPE simulations without any adjustable parameters. 
For $t_\Omega < 200\,\si{ms}$, $\mathcal{N}_v$ is below 1, where vortices, if present, are at the edge of the cloud. For longer times, $\mathcal{N}_v$ increases and saturates to an average value of about three and a maximum of six vortices (see Fig.\,\ref{fig:Fig3}a for an example of five vortices). The observed saturation might be due to the decreased visibility and to the atom-loss-induced shrinking of the BEC size, which is not accounted for in the theory. We  also compare the course of the average vortex number with the AR of the cloud. After initial large oscillations, due to the sudden jump in rotation frequency, the AR declines towards  $\approx$ 1 \cite{Madison2001sso}. This happens as the vortex number simultaneously increases. 

\begin{figure*}[tb!]
\includegraphics[width=6.72 in, ]{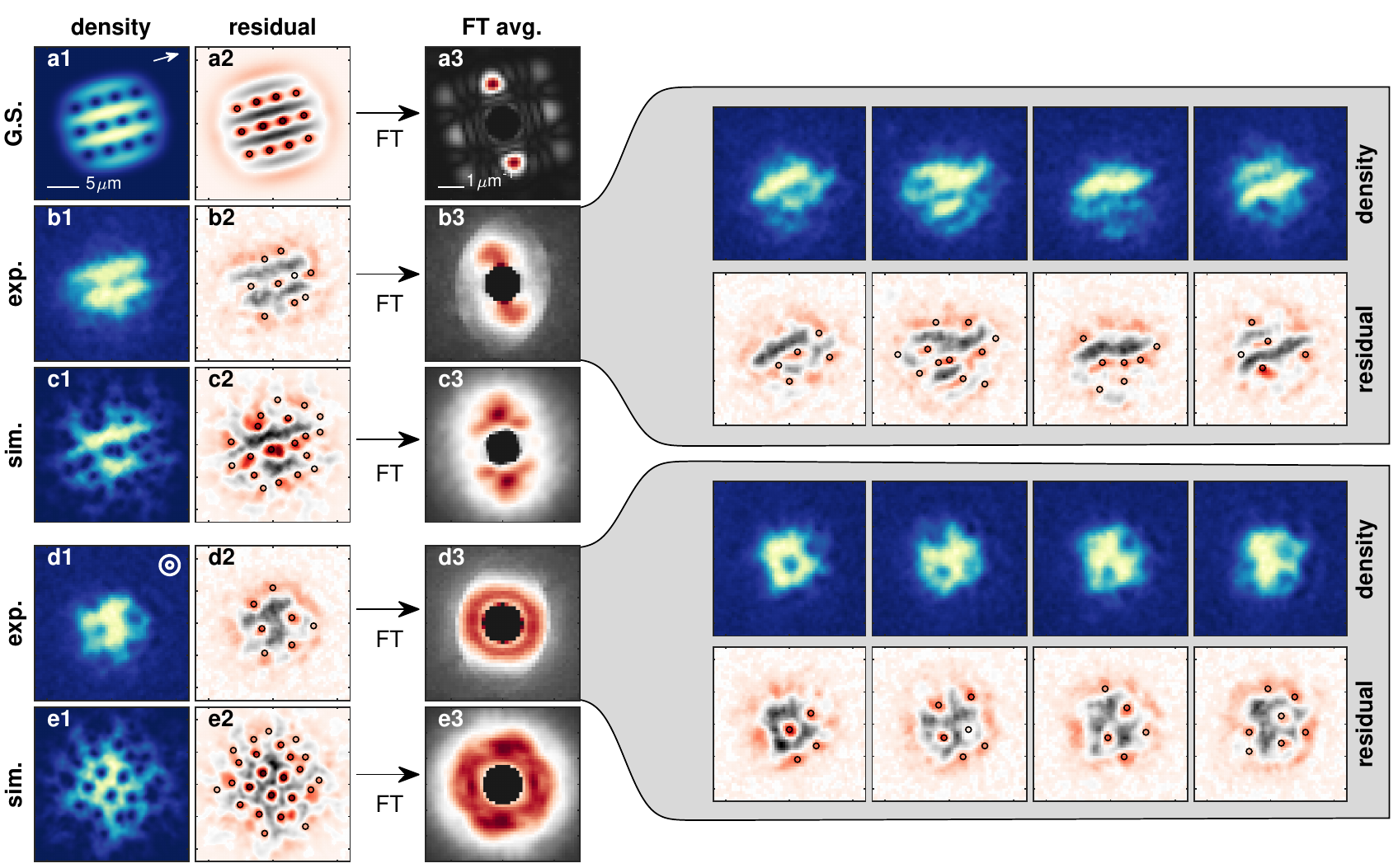}
\caption{
\textbf{Stripe nature of dipolar vortices.} \textbf{a1}, Ground state stripe lattice solution for our experimental parameters $a_s = 109\, a_0$, trap frequencies $(\omega_{\perp}, \omega_z) = 2\pi \times [50,130]\,\si{Hz}$, $N$=10000 and $\Omega = 0.75\,\omega_\perp$. \textbf{a2}, Corresponding residual image, found by subtracting the ground state from the blurred image, with circles showing the detected vortices. \textbf{a3}, Fourier transform of the residual image. \textbf{b1}, Single experimental image after $500\,\si{ms}$ of continuous rotation at $\Omega=0.75\,\omega_\perp$. \textbf{b2}, The corresponding residual image. \textbf{b3}, Fourier transform of the residual images, averaged over 49 runs, with example shots shown to the right.
\textbf{c1,} Simulation result for the dynamic experimental procedure in \textbf{b}.  \textbf{c2-c3}, Residuals and FT analysis (115 temporal images) as in \textbf{b2-b3}. \textbf{d1-d3}, Same as \textbf{b1-b3} for 121 runs, but we rotate for an additional $100\,\si{ms}$ and then spiral the magnetic field to $\theta=0^\circ$ over a further $100\,\si{ms}$ before imaging. \textbf{e1-e3}, Simulation result for procedure in \textbf{d}. All simulation images are rotated to have the same magnetic field direction as the experiment.}
\label{fig:stripes}
\end{figure*}

One fascinating prediction with dipolar vortices under the influence of a rotating magnetic field relates to the structure of the resulting vortex lattice. Due to magnetostriction and the anisotropic vortex cores, the resulting vortex configuration is also anisotropic, producing a stripe phase in the strongly dipolar regime \cite{yi2006vsi,Martin2017vav}, instead of the usual triangular lattice in non-dipolar BECs \cite{Madison2000vfi}. The ground state stripe lattice solution for our parameters is shown in Fig.\,\ref{fig:stripes}a, with a cloud ${\text{AR} = 1.08}$. In the vortex stripe phase, vertical planes of high density regions, parallel to the magnetic field, alternate with low density ones, that host rows of vertical vortex filaments. Such a configuration promotes head-to-tail dipolar attraction within the high density ridges, and this acts to lower the energy. It should be noted that these states are distinct from the oscillating vortex sheets states, which appear after squeezing a triangular vortex lattice \cite{Engels2002neo}.

To explore this prediction, we perform two new surveys. First, we slightly reduce the magnetic field value, reducing the scattering length to $a_s\approx109\,a_0$ and hence making the system relatively more dipolar, while still avoiding the droplet regime. We magnetostir the BEC at a constant rotation frequency $\Omega = 0.75\,\omega_\perp$ for $500\,\si{ms}$, but during TOF we stop the magnetic-field rotation and keep it in place at $\theta=35^\circ$. The stripe structure is revealed in Fig.\,\ref{fig:stripes}b1 for a single experimental run, and is clearly visible in the residual image, Fig.\,\ref{fig:stripes}b2, where the vortices align along three stripes. The spatial structure of the residual image can be assessed through the absolute value of two-dimensional Fourier transform (FT). After taking the FT of each residual image, we then average the result, see Fig.\,\ref{fig:stripes}b3, finding a clear peak at the $k$ of the inter-stripe spacing. This shows that the stripe spatial structure survives the averaging, implying that the majority of images show stripes with the same spacing, and they also have the same orientation as set by the magnetic field, as evidenced by the example images shown in the right of Fig.\,\ref{fig:stripes}.
Note that these observations do not rely on our ability to resolve individual vortices, as the stripes are an ensemble effect of many aligned vortices. In fact, by comparing with the numerical simulations of the dynamical procedure (Fig.\,\ref{fig:stripes}c1-c3), we expect there are more vortices than detected here that fill in the stripes, forging out this structure. In general, our simulations show that the stripes appear faster when the scattering length is lower, and when the atom number is larger. In the long time limit of the scenario presented in Fig.\,\ref{fig:Fig2}, we expect the stationary solution to also be the stripe state, but this is not observable on our timescales.

Remarkably, the stripe structure washes out when we subsequently tilt the magnetic-field orientation to $\theta=0^\circ$ (parallel to the trap symmetry axis), as shown in Fig.\,\ref{fig:stripes}d1. Here, after $600\,\si{ms}$ of magnetostirring, we add another step in which we spiral up the magnetic field to ${\theta=0^\circ}$ (with $\Omega$ fixed) over $100\,\si{ms}$, before imaging. Under these conditions, all vortex properties are again isotropic within the plane. The final non-equilibrium positioning of the vortices is arbitrary,
and if we average the FT of the residuals directly, we observe a homogeneous ring in the average FT (Fig.\,\ref{fig:stripes}d3). Also, this behavior is confirmed by the simulations, as shown in Fig.\,\ref{fig:stripes}e1-e3.

By exploiting magnetostirring --- a novel, robust, method of generating angular momentum --- we have observed quantized vortices in a dipolar quantum gas, and the appearance of the vortex stripe configuration. Future works will focus on investigations of the individual vortex shape and behaviour, such as the anisotropic nature of the vortex cores for in-plane magnetic fields \cite{yi2006vsi,Ticknor2011asi,mulkerin2013anisotropic,mulkerin2014vortices}, the interplay between the vortex and roton excitations \cite{yi2006vsi,Ticknor2011asi,mulkerin2013anisotropic,mulkerin2014vortices,JonaLasinio2013rci}, exotic vortex patterns such as square lattices \cite{Martin2017vav}, and investigations into anisotropic turbulence \cite{Bland2018qft}. This work also opens the door to studying more complex matter under rotation, such as dipolar droplets \cite{Cidrim2018vis,Lee2018eoa,FerrierBarbut2018smo} and supersolid states \cite{Gallemi2020qvi,Roccuzzo2020ras,Ancilotto2021vpi,Norcia2021cao}. Such proposals will be challenging due to the intricate density patterns \cite{Hertkorn2021pfi}, however such observations would provide conclusive evidence of superfluidity in supersolids. Rotating the magnetic field at frequencies far larger than the radial trap frequencies, but smaller than the Larmor frequency, has been observed to tune the sign and magnitude of the dipole-dipole interaction \cite{Tang2018ttd, Giovanazzi2002ttd}--a method also employed in Nuclear Magnetic Resonance spectroscopy \cite{Maricq1979nir}--but there remain open questions on the stability of this procedure \cite{Prasad2019ior,Baillie2020rto}, which if rectifiable would unlock new research directions \cite{Tang2018ttd}. Other vortex generation methods, such as thermally activated pairs in quasi-2D to assess the Berezinskii-Kosterlitz-Thouless transition, and stochastically generated vortex tangles through temperature quenches to assess the Kibble-Zurek mechanism, remain unexplored in dipolar gases \cite{Martin2017vav}. The technique introduced here is also applicable to a wide range of systems governed by long-range interactions through the manipulation of magnetic or electric fields.\\

\textbf{Acknowledgements:} \\
We are grateful to S.\,B.\,Prasad, M.\,Norcia, R.\,M.\,W.\,van Bijnen, and L. Santos for helpful discussions.
We acknowledge M.\,Norcia and A.\,Patscheider for experimental contributions. 
This study received support from the European Research Council through the Consolidator Grant RARE (No.\,681432), the QuantERA grant MAQS by the Austrian Science Fund FWF No.\,I4391-N, the DFG/FWF via FOR 2247/PI2790 a joint-project grant from the FWF (Grant No.
I4426, RSF/Russland 2019).
M.J.M.\,acknowledges support through an ESQ Discovery Grant by the Austrian Academy of Sciences. We also acknowledge the Innsbruck Laser Core Facility, financed by the Austrian Federal Ministry of Science, Research and Economy. Part of the computational results presented have been achieved using the HPC infrastructure LEO of the University of Innsbruck.
G.L.\,acknowledges financial support from Provincia Autonoma di Trento. \\

\textbf{Author contributions:} \\
L.K., C.P., G.L., E.C., M.J.M., and F.F.\,performed the experimental work and data analysis. E.P., T.B., and R.B.\,performed the theoretical work. All authors contributed to the interpretation of the results and the preparation of the  manuscript.

\appendix

\newpage

\section{Methods}
\renewcommand{\figurename}{Extended Data Fig.}
\setcounter{figure}{0}

\subsection{Experimental Procedure}

We prepare an ultracold gas of $^{162}$Dy atoms in an optical dipole trap (ODT). Three 1064 nm laser beams, overlapping at their foci, form the ODT. The experimental procedure to Bose-Einstein condensation is similar to the one followed in our previous work \cite{Norcia2021cao}, but the magnetic field unit vector, $\vu*{\rm B}$, is tilted by an angle of $\theta = 35$° with respect to the $z$-axis during the whole sequence. After preparation, the sample contains about $2 \times 10^4$ condensed atoms. The corresponding trap frequencies are typically $(\omega_\perp, \omega_z) = 2\pi\times[50.8(2), 140(1)]\,\si{Hz}$. For all our measurements, the deviation of the trap aspect ratio in the $xy$-plane AR$_\text{trap}=\omega_y/\omega_x$ from 1 is always smaller than $0.6\%$.
We evaporate the atoms at $|{\va*{\rm B}}| = 5.423(5)\,\si{G}$ and jump to the final magnetic-field value during the last evaporation ramp. After the preparation of the BEC, the magnetic field is rotated as described in the next section. 
We use standard absorption imaging to record the atomic distribution. We probe the vortices using the vertical imaging taken along the axis of rotation ($z$), for which the dark spots within the condensate correspond to the cores of individual vortices. The vertical images are taken with a short TOF of $3\,\si{ms}$ and a pulse duration of 3-4\,$\mu$s. 
For the  data in Fig.\,1-3, we let the magnetic field  spinning during TOF, whereas for Fig.\,4 we use a static field orientation.

\subsection{Control of the magnetic field}

{\em Calibration}: Three pairs of coils -- each oriented along a primary axis in the laboratory frame -- enable the creation of a homogeneous field with arbitrary orientations.  The absolute magnetic field value $|\va*{\rm B}|$ of each pair of coils is independently calibrated using radio frequency (RF) spectroscopy. The RF drives transitions to excited Zeeman states, leading to a resonant dip in the atom number. The long-term stability -- measured via the peak position of the RF resonance over the course of several days -- is on the order of $\Delta B = \pm1\,\si{mG}$, while shot-to-shot fluctuations -- measured via the width of the RF resonance for a single calibration set -- is $\Delta B = \pm5\,\si{mG}$. \\
{\em Rotation}: We drive the rotation of the magnetic field by sinusoidally modulating the magnetic-field value components $B_x$ and $B_y$ with a phase difference of $90^\circ$ between them. Since we want to keep the absolute magnetic field value $|{\va*{\rm B}}|$ constant during rotation, we measure it for various values of the azimuthal angle $\phi$ and fixed $\theta = 35^\circ$ by performing Feshbach loss spectroscopy around $5.1\,\si{G}$. We find an average shift of $|{\va*{\rm B}}|$ of about $10\,\si{mG}$ from the $\theta=0^\circ$ case, which we take into account. We also find small deviations as a function of $\phi$ of $\Delta |{\va*{\rm B}}| < 20\,\si{mG}$, which might appear due to slightly non-orthogonal alignment of the magnetic fields. We did not correct these deviations for the sake of simplicity. 

\subsection{Scattering length}
The scattering length in $^{162}$Dy is currently not known with large accuracy \cite{Tang2015sws,Tang2016aeo,Lucioni2018ddb,Boettcher2019ddq}. To estimate the scattering length in the small magnetic-field range  around $B = 5.3\,\si{G}$, relevant to this work, we use the well-known relation $a_s =  a_{\text{bg}} \prod_i [1 - \Delta B_i/ (B - B_{0,i})]$ \cite{Chin2010fri}, where $B_{0,i}$ and $\Delta B_i$ are the center position and the width of the Feshbach loss features reported in Ref.\,\cite{Boettcher2019ddq}, respectively. 

\begin{figure}[h!]
    \includegraphics[width=3.38 in, ]{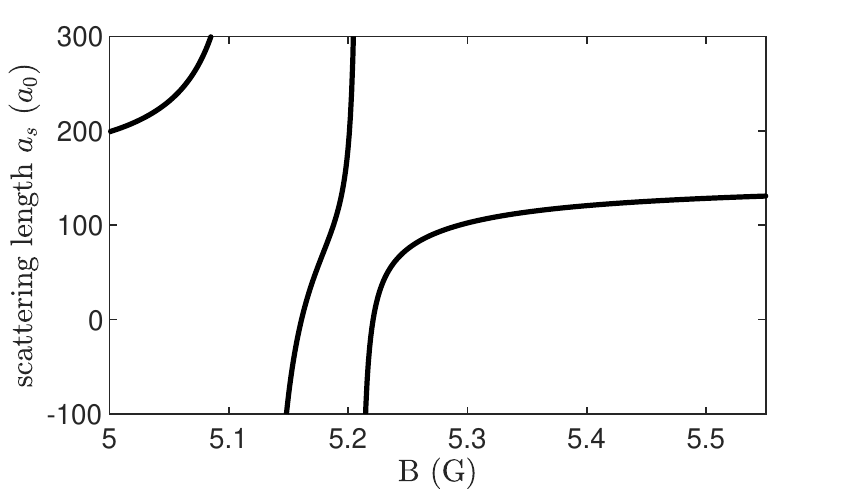}
    \caption{\textbf{Calculated ${\rm B}$-to-$a_s$ conversion for $^{162}$Dy.} Scattering length as a function of the magnetic-field value with the background scattering length  $a_{\text{bg}} = 129(9)\,\si{a_0}$. We find $a_s = 111(9)\,\si{a_0}$ at $B = 5.333\,\si{G}$.}
    \label{fig:extfig_BVsas}
\end{figure}

The value of the background scattering length, $a_{\text{bg}}$, is empirically fixed by measuring the magnetic field value at which the supersolid transition occurs and comparing it with the corresponding critical $a_s$ predicted from simulations.  
Such an approach leads to $a_s= 111(9)\,\si{a_0}$ at $B = 5.333\,\si{G}$. Extended Data Fig.\,\ref{fig:extfig_BVsas} shows the resulting scattering lengths for the relevant magnetic fields.
Although such an approach gives very good agreement between theory and experiments, future works on a precise determination of $a_s$, similar to the one achieved with erbium \cite{Patscheider2022dot}, would be desirable. 

\subsection{Magnetostirring}
Tilting the magnetic field vector ${\va*{\rm B}}$ away from the symmetry axis of our cylindrical trap leads to an ellipsoidal deformation of the cloud \cite{Stuhler2005ood} and therefore to a breaking of the cylindrical symmetry. This allows for the transfer of angular momentum to the sample by rotating the magnetic field (magnetostirring). In all our measurements we use a ${\va*{\rm B}}$ tilted with respect to the $z$-axis by 35° and a constant value $|{\va*{\rm B}}|$. That value is $|{\va*{\rm B}}| = 5.333(5)\,\si{G}$ for the surveys in Figs.\,\ref{fig:Fig1}-\ref{fig:Fig3} and $|{\va*{\rm B}}| = 5.323(5)\,\si{G}$ for Fig.\,\ref{fig:stripes}. For these parameters, the magnetostricted aspect ratio of the cloud is $\mathrm{AR} = 1.03$. For all our measurements, the measured trap $\mathrm{AR}_\text{trap} < 1.006$ is much smaller than the deformation due to magnetostriction. Additionally, we have confirmed with simulations that even with trap asymmetries of up to 10\%, e.g.\,$(\omega_x,\omega_y)=(55,50)$Hz, this procedure can still generate vortices in a lattice configuration.  \\
{\em Adiabatic frequency ramp}: We employ different magnetic-field rotation sequences for the different data sets. For the data set of Fig.\,\ref{fig:Fig1}c, the rotation frequency of the magnetic field is linearly increased to different final values at a speed of $\dot{\Omega} = 2\pi \times 50\, \si{Hz}/\si{s}$ and for a duration of ${t_{\dot{\Omega}} = 0\text{-}1\,\si{s}}$. The ramp time is much longer than the period of the rotation $\Omega^{-1}$ for higher rotation frequencies $\Omega \gtrsim \Omega_c$, and therefore the ramp is adiabatic for the regimes considered, until the onset of dynamical instabilities. After the ramp, the magnetic field direction is rotated at the target rotation frequency $\Omega$ for one final period (as shown in Fig.\,\ref{fig:Fig1}b2). We sample 10 different final magnetic field angles during this last rotation, measuring the corresponding aspect ratio and averaging the result in order to remove any potential biases due to latent trap anisotropies. Each data point is then obtained with 8-10 experimental runs. \\
{\em Constant rotation frequency}: For the dataset of Fig.\,\ref{fig:Fig2}, Fig.\,\ref{fig:Fig3} and Fig.\,\ref{fig:stripes}b we directly start to rotate at the final rotation frequency $\Omega$ without any acceleration phase. The magnetic field is then rotated for a variable time $t_\Omega$, after which the atoms are released from the trap and a vertical image is taken. \\
{\em Spiral up magnetic field}: For the dataset of Fig.\,\ref{fig:stripes}d we employ a similar sequence as described above. However, after constantly rotating the magnetic field at $\Omega=0.75\,\omega_\perp$, the magnetic field is spiraled up in $100\,\si{ms}$ to $\theta = 0$° by linearly reducing $\theta$ while continuing rotating. Afterwards, the atoms are released from the trap and a vertical image is taken.

\subsection{Theoretical model}
We employ an extended Gross-Pitaevskii formalism to model our experimental setup. In this scheme, the inter-particle interactions are described by the two-body pseudo-potential, 
\begin{align}
    U(\textbf{r}) = \frac{4\pi\hbar^2a_{\rm s}}{m}\delta(\textbf{r})+\frac{3\hbar^2a_\text{dd}}{m}\frac{1-3\left(\hat{\mathbf{e}}(t)\cdot\mathbf{r}\right)^2}{r^3}\,,
\end{align}
with the first term describing short-range interactions governed by the s-wave scattering length $a_s$, with Planck's constant $\hbar$ and particle mass $m$. The second term represents the anisotropic and long-ranged dipole-dipole interactions, characterized by dipole length $a_\text{dd}=\mu_0\mu_m^2m/12\pi\hbar^2$, with magnetic moment $\mu_m$ and vacuum permeability $\mu_0$. We always consider $^{162}$Dy, such that $a_\text{dd}=129.2\,a_0$, where $a_0$ is the Bohr radius. The dipoles are polarized uniformly along a time-dependent axis, given by
\begin{align}
    \hat{\mathbf{e}}(t)=(\sin\theta(t)\cos\phi(t),\sin\theta(t)\sin\phi(t),\cos\theta(t))\,
\end{align}
with time dependent polarization angle $\theta(t)$ and $\phi(t)=\int_0^t \text{d}t'\Omega(t')$, for rotation frequency protocol $\Omega(t)$.

Beyond-mean-field effects are treated through the inclusion of a Lee–Huang–Yang correction term \cite{Lima2011qfi}
\begin{align}
    \gamma_\text{QF}=\frac{128\hbar^2}{3m}\sqrt{\pi a_s^5}\,\text{Re}\left\{ \mathcal{Q}_5(\edd) \right\} \, ,
\end{align}
with $\mathcal{Q}_5(\edd)=\int_0^1 \text{d}u\,(1-\edd+3u^2\edd)^{5/2}$ being the auxiliary function, and the relative dipole strength is given by $\edd = a_{\rm dd}/a_{\rm s}$. Finally, the full extended Gross-Pitaevskii equation (eGPE) then reads \cite{Waechtler2016qfi,FerrierBarbut2016ooq, Chomaz2016qfd,Bisset2016gsp}
\begin{align}
    &i\hbar\frac{\partial\Psi(\textbf{x},t)}{\partial t} =  \bigg[-\frac{\hbar^2\nabla^2}{2m}+\frac12m\left(\omega_x^2x^2+\omega_y^2y^2+\omega_z^2z^2\right)  \nonumber\\
    &+ \int\text{d}^3\textbf{x}'\, U(\textbf{x}-\textbf{x}')|\Psi(\textbf{x}',t)|^2  +\gamma_\text{QF}|\Psi(\textbf{x},t)|^3\bigg]\Psi(\textbf{x},t)\,,
    \label{eqn:GPE}
\end{align}
where $\omega_{x,y,z}$ are the harmonic trap frequencies. The wavefunction $\Psi$ is normalized to the total atom number $N=\int {\rm d}^3\mathbf{x}|\Psi|^2$. The stationary solution for Fig.\,\ref{fig:stripes}a from the main text is found through the imaginary time procedure in the rotating frame, introducing the usual angular momentum operator $\Omega\hat{L}_z$ into Eq.\,\eqref{eqn:GPE}.
The initial state $\Psi(\textbf{x},0)$ of the real-time simulations is always obtained by adding non-interacting noise to the ground state $\Psi_0(\textbf{x})$. Given the single-particle eigenstates $\phi_n$ and the complex Gaussian random variables $\alpha_n$ sampled with $\langle|\alpha_n|^2\rangle = (e^{\epsilon_n/k_BT}-1)^{-1}+\frac12$ for a temperature $T=20\,\si{nK}$, the initial state can be described as $\Psi(\textbf{x},0) = \Psi_0(\textbf{x}) + \sum_n' \alpha_n \phi_n(\textbf{x})$,  where the sum is restricted only to the modes with $\epsilon_n\le2k_BT$ \cite{blakie2008das}. Throughout, the density images are presented in situ, with a scaling factor to account for the 3ms TOF for the experimental images.

In order to obtain the average residual Fourier Transform images for Figs.\,\ref{fig:stripes}c3 and e3, we first Fourier Transform 115 frames from the simulation between 700ms and 1.1s in the rotating frame before averaging the result.

\subsection{Atom number}

\begin{figure}[h!]
    \centering
    \includegraphics[width=3.38 in,]{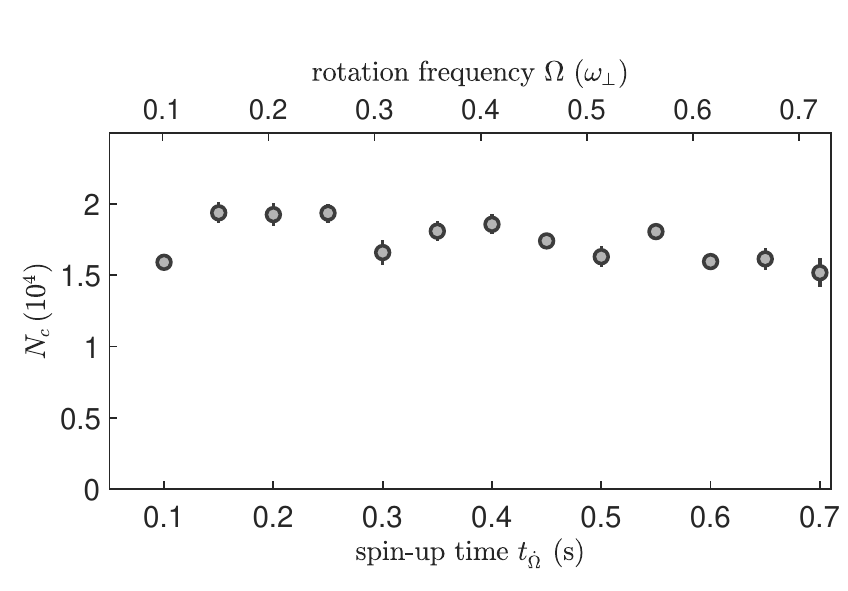}
    \caption{\textbf{Condensed atom number $N_c$  during magnetostirring (Fig.\,1c)}. Condensed atom number as a function of spin-up time $t_{\dot{\Omega}}$ for the same sequence as in Fig.\,\ref{fig:Fig1}c. The condensed atom number is extracted by fitting a two-dimensional bimodal distribution of Thomas-Fermi and Gaussian  function to the horizontal density distributions.} 
    
    \label{fig:fig1_atom_number}
\end{figure}

Extended Data Fig.\,\ref{fig:fig1_atom_number} shows the condensed atom number $N_c$ for the measurement with an adiabatic ramp of the magnetic-field rotational velocity ($\dot{\Omega} = 2\pi \times 50\,\si{Hz}/\si{s}$), corresponding to the data of Fig.\ref{fig:Fig1}c of the main text. To extract the atom number, we use the horizontal imaging with $26\,\si{ms}$ of TOF. About $3\,\si{ms}$ before flashing the imaging resonant light to the atoms, we rotate the magnetic field in the imaging plane and perform standard absorption imaging. From the absorption images, we extract $N_c$ from a bimodal fit up to $700\,\si{ms}$. At later times, the system undergoes a dynamic instability (see discussion in the main text) and the density profile deviates from a simple bimodal distribution. During the observation time, we see a slight decrease of $N_c$ and for our theory simulations we use a constant atom number of $N_c=15000$. Note that in all following datasets, in which we abruptly accelerate the magnetic-field rotation to the desired final velocity, we observe a faster decay and our simulations are performed with either $N_c=8000$ or $N_c=10000$.

\subsection{Vortex detection}

{\em Vortex detection algorithm:} Since vortices appear as dark holes in the density profile of a BEC, which would otherwise have a smooth profile, our approach to extract the number of vortices is to look at deviations between the image and an unmodulated reference image. To extract the vortex number from the raw images, we proceed as follows: 

First, we prepare the image $n_{\text{img}}$, the reference image $n_{\text{ref}}$ and the residual image $n_{\text{res}}$. The image is normalized such that the maximum density $\max(n_{\text{img}})=1$. We create the reference image by blurring the image via applying a 2D Gaussian filter with $\sigma = 5\,$pixel, corresponding to about $2.1\,\mu$m. This blurring smoothens any structure on the lengthscale of the filter width, therefore any holes in the density profile wash out. We then normalize the atom number of the reference to be the same as from the image $N_{\text{ref}} = \int\int n_{\text{ref}} \doteq N_{\text{img}} = \int\int n_{\text{img}}$. The residual image is calculated as the difference between the image and the reference $n_{\text{res}} = n_{\text{img}}-n_{\text{ref}}$. We additionally mask the region where the density of the reference is below a certain threshold ($n_{\text{res}}=0$ where $n_{\text{ref}}\leq 0.1$).

Second, we identify local minima in the residual image and determine if they are connected to vortices. For this, we create a list of local minima $(x_{\text{min}},y_{\text{min}})$, defined by the condition that the pixel density $n_{\text{res}}(x_{\text{min}},y_{\text{min}})$ is lower than of all surrounding pixels. Then we remove minima with density values above zero $n_{\text{res}}(x_{\text{min}},y_{\text{min}})\geq 0$ or which are within one pixel distance of the mask border. Now we determine a local contrast for each minimum by calculating the difference between its central density value and the mean of the density values $\pm2\,$ pixel values away from it $n_{\text{con}}(x_{\text{min}},y_{\text{min}})=n_{\text{res}}(x_{\text{min}},y_{\text{min}})-\text{mean}( n_{\text{res}}(x_{\text{min}}\pm2\,$px$,y_{\text{min}}\pm2\,$px$))$, and remove minima above a certain threshold $n_{\text{con}}>-0.11$. As a last step we check the distance $d$ between all remaining minima to avoid double counting of minima too close to each other. In case $d$ is below the threshold $d<5\,$px the minimum with the higher residual density value $n_{\text{res}}$ is discarded. 

{\em Preparation of theory density profiles:} For the direct comparison of the vortex number shown in Fig.~\ref{fig:Fig3}b, we apply additional steps to the density profiles obtained from theory. First, we reduce the resolution by a $2\times2$ binning to make the pixel size of the theory density profiles $n^\text{theo}_{\text{img}}$ essentially the same as for the experimental images (sizes are within $5\%$). After normalizing to $\max(n^\text{theo}_{\text{img}})=1$ we apply Gaussian white noise with zero mean and a variance of $0.01$ to each pixel, recreating the noise pattern from empty regions of experimental images. Then we blur the image using a 2D Gaussian filter with $\sigma = 1\,$pixel ($\sim0.42\,\mu$m), this recreates the same resolution condition as our experimental setup. The resulting density profile is taken as the input image for the vortex detection algorithm described above.

{\em Benchmarking the vortex detection algorithm:} As the vortex positions for the simulation images are known a priori due to the available phase map, we can derive the fidelity of the vortex detection algorithm for simulation data. For the theory data shown in Fig.~\ref{fig:Fig3}b in the time frame between $600$ and $700\,\si{ms}$, the average detected vortex number in the simulated density profiles (applying the preparation scheme described above) is about 9, while the real number of vortices present in the same area of the image is about 33 in average. This mismatch is explained by the conservative choice of the thresholds for vortex detection together with the added noise, which results in only counting clear density dips as vortices, throwing out many vortices in the low density region. This conservative choice of thresholds on the other hand leads to a very high fidelity of $>97\%$, where we define the fidelity as the percentage of detected vortices which correspond to actual present vortices in the data. For raw simulation data (without resolution reduction, added noise and blurring) the vortex detection algorithm would detect up to $80\%$ of the vortices present with a fidelity of $>95\%$.

Note that for the visualization of the vortex positions for Fig.~\ref{fig:stripes} we slightly increased the local threshold $n_{\text{con}}>-0.08$ and decreased the minimum distance between vortices $d<3\,$px, which increases the overall number of vortices detected. For the density distributions obtained from theory, we additionally omit the resolution reduction, addition of noise and blurring steps.

\newpage

\section{Extended data figures}

\newpage

\begin{figure*}[htb!]
    \centering
    \includegraphics[width=2\columnwidth ]{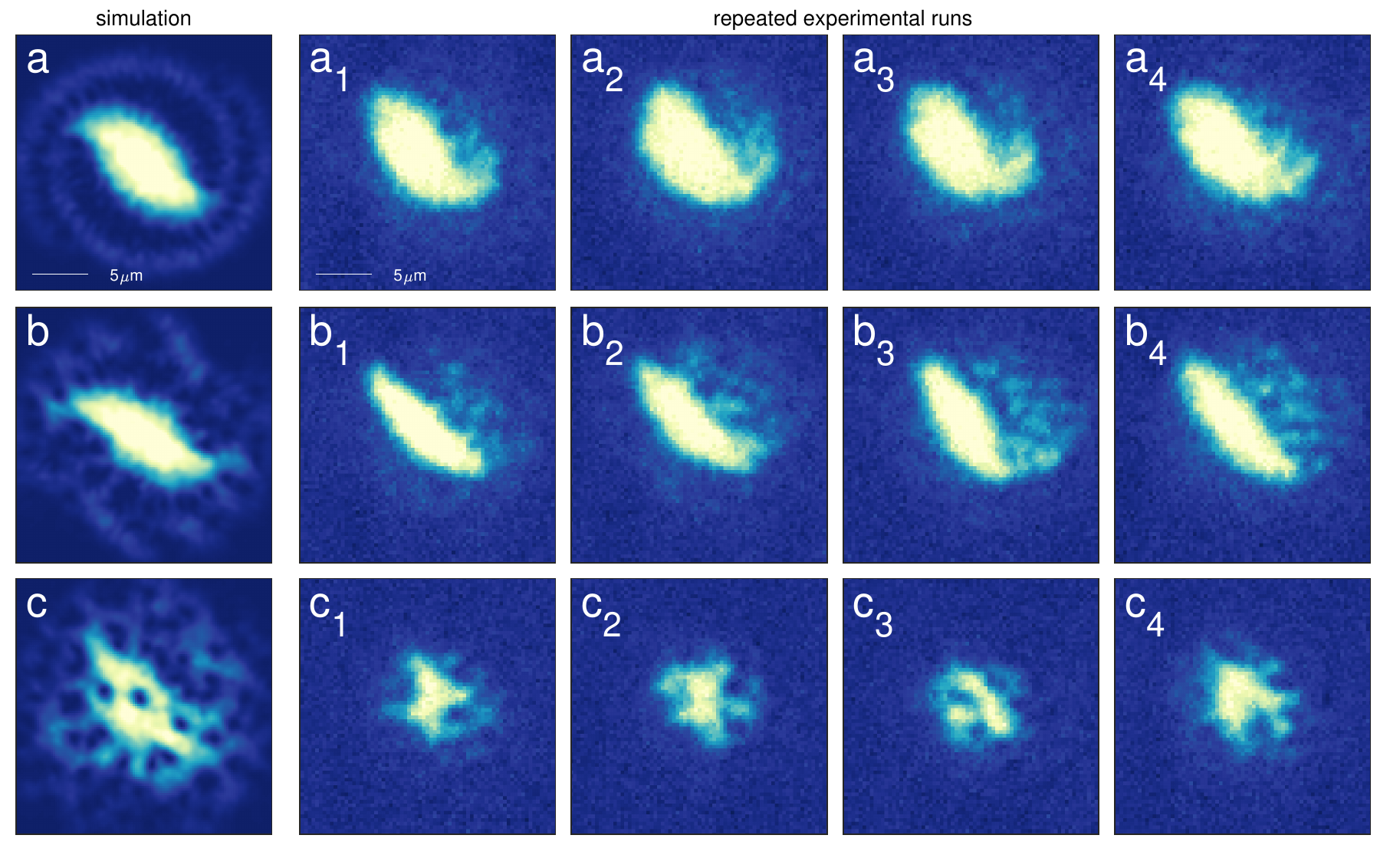}
    \caption{\textbf{Repeatability of the vortex generation protocol.} Each row shows the simulated image (\textbf{a, b, c})  and the corresponding vertical TOF images from independent experimental runs ({\boldmath$\mathrm{a}_i$, $\mathrm{b}_i$, $\mathrm{c}_i$}) for a different rotation time: $t_a=127\, \si{ms}$, $t_b=207\, \si{ms}$, and $t_c=741 \,\si{ms}$. The rotation frequency is $\Omega=0.74 \,\omega_\perp$ with the trap frequencies being $\omega_t=2\pi\times[50.7(1),50.8(1),129(1)]\,\si{Hz}$. The magnetic field value is $|{\va*{\rm B}}| =5.333(5)\,\si{G}$. 
    For the simulation the scattering length used is $112\,\si{a_0}$, the trap frequencies are $(50,50,150)\,\si{Hz}$, the condensed atom number is $N=8000$, and the rotation frequency is $0.75\,\omega_\perp$.
    }
    \label{fig:fig2_repeat_exp}
\end{figure*}

\end{document}